\documentclass[a4paper]{article}
\author{}

\usepackage[
  backend=bibtex,
  style=numeric,
  sorting=none,
  maxnames=3,
  minnames=3,
  giveninits=true
]{biblatex}

\usepackage{geometry}
\usepackage{multirow}
\usepackage{dblfloatfix}
\usepackage{float}
\usepackage{placeins}
\usepackage{graphicx, hyperref, setspace, amsmath, amssymb, titlesec, fancyhdr, multicol, parskip, indentfirst, etoolbox, caption, 
hyperref, xcolor, booktabs}

\titleformat{\section}{\centering\large\scshape}{\thesection}{1em}{}
\titleformat{\subsection}{\normalsize\bfseries}{\thesubsection.}{1em}{}
\titlespacing{\section}{0pt}{6pt}{6pt}
\titlespacing{\subsection}{0pt}{6pt}{6pt}
\titlespacing{\subsubsection}{0pt}{6pt}{6pt}
\title{
    \textbf{Detector-aware target definitions for full-event particle reconstruction} 
}

\date{} 
\renewcommand{\thesection}{\arabic{section}}

\titleformat{\subsection}{\normalfont\large\itshape}{\thesubsection}{1em}{}
\titleformat{\subsubsection}{\normalfont\itshape}{\thesubsubsection}{1em}{}

\begin{document}

\maketitle
\vspace{-1.5cm}

\begin{center}
\textbf{Katharina Schäuble$^1$, Alessandro Brusamolino$^1$, Dolores Garcia$^2$, Jan Kieseler$^1$} \\
\textit{$^1$Institut für Experimentelle Teilchenphysik, Karlsruher Institut für Technologie, Karlsruhe, Germany}\\
\textit{$^2$CERN, Meyrin, Switzerland}\\
\textit{Email: \href{mailto:katharina.schaeuble@cern.ch}{\texttt{katharina.schaeuble@cern.ch}}, \href{mailto:alessandro.brusamolino@cern.ch}{\texttt{alessandro.brusamolino@cern.ch}}, \href{mailto:dolores.garcia@cern.ch}{\texttt{dolores.garcia@cern.ch}}, \href{mailto:jan.kieseler@cern.ch}{\texttt{jan.kieseler@cern.ch}}}
\end{center}

\singlespacing
\setlength{\parskip}{6pt}
\setlength{\parindent}{0.5cm}

\begin{abstract}
Hit-level ML-based particle reconstruction methods have recently shown promising results.
However, the reconstruction models are currently provided with targets that are unaware of the detector geometry and its resolution, resulting in training ambiguities. This can introduce a dependence on sample priors and reduce robustness under changes in event topology.
We study the effect of a detector-aware target definition in the context of end-to-end Particle Flow reconstruction using a generic GEANT4-based detector simulation. 
We introduce the concept of detector-aware targets built from calorimeter showers with a hit-based merging algorithm based on cell-wise energy sharing that takes into account the spatial resolution of the detector. 
This includes a Particle-Flow-aware variant that preserves charged-particle consistency.
Using a fixed GNN-based reconstruction model, we show that merged targets improve physics performance on a training-like sample. More importantly, models evaluated on an independent sample with different particle composition and topology show improved momentum response and resolution when trained with PF-aware merged targets. Our results show that removing experimentally non-resolvable target structure enhances not only reconstruction performance, but also improves model robustness against process-dependent variations in event topology.
\end{abstract}

\begin{multicols}{2}
\setlength{\columnsep}{0.5cm}

\section{Introduction}
High Energy Physics analyses rely on reconstructed representations of the stable particles produced in a collision. These representations are obtained through a sequence of reconstruction steps, starting from local reconstruction within individual subdetectors and progressing toward global event interpretation. Particle Flow algorithms form one such global reconstruction stage: they combine information from tracking, calorimetry, and other subdetectors to reconstruct particle candidates and estimate their properties, such as energy, momentum, and direction \cite{PF_Sirunyan_2017, PF_ATLAS_2017, PandoraPFA}. Recent ML-based approaches, such as Ref.~\cite{MLPF_Pata, MLPF, garcia2026end}, are posed as supervised learning tasks in which the targets are defined as low-level detector information, thereby making the definition of the reconstruction target itself an even more central part of the reconstruction problem.

The choice of reconstruction target determines both the physics objects that an algorithm is trained or designed to recover and how its performance is evaluated. However, this choice is not unique: Ref.~\cite{MLPF,MLPF_Pata} defines the targets using particles that are stable at generator level (status 0), whereas Ref.~\cite{kakati2025hgpflowextendinghypergraphparticle} additionally includes some particles with Pythia generator status 2 and particles produced through interaction with the detector (status 1). Ref.~\cite{garcia2026end} takes an intermediate approach using the calorimeter boundary as an abstraction layer for the definition. However, none of these works include detector resolution in the definition. 
In particular, finite detector resolution can make distinct particle configurations experimentally indistinguishable. For example, two particles entering the calorimeter with a separation comparable to or smaller than the effective cell granularity may produce overlapping energy deposits that cannot be separated from the detector response alone, as depicted in Fig.~\ref{fig:shower_overlap}. In such cases, treating the two particles as separate reconstruction targets introduces an ambiguity that no reconstruction algorithm can resolve deterministically. The target definition must therefore encode the experimentally resolvable structure of the detector response, rather than only the more granular information available from simulation.

If the same detector-level information is compatible with several simulation-level target decompositions, the reconstruction model cannot learn a unique mapping from observables to targets. Instead, it must resolve the ambiguity through correlations present in the training sample, thereby introducing a dependence on sample priors. In this paper, we study this problem in the context of Particle Flow reconstruction, focusing on ambiguities arising from overlapping calorimeter showers. We construct detector-aware training targets through a hit-based shower merging procedure, while preserving the cross-subdetector consistency required for Particle Flow. The procedure does not require explicit geometric distance criteria, with detector granularity entering solely through the pattern of shared detector hits.
The hypothesis is that removing non-resolvable target structure improves not only reconstruction performance on the training-like sample, but also the model's robustness when applied to samples with different particle composition and detector topology.

The paper is structured as follows. Section~2 describes the simulated detector setup. Section~3 defines the simulation-level shower objects used as the starting point for target construction. Section~4 introduces the hit-based merging algorithm and its global and Particle-Flow-motivated iterative variants. Section~5 validates the effect of the merging procedure on the resulting target objects. Section~6 studies the impact of the modified targets on ML-based Particle Flow reconstruction, including their effect on robustness under a change in jet topology.

\section{Detector simulation and reconstructed objects}\label{sec:sim_and_reco}

The samples used in this paper have been generated using a GEANT4~\cite{Allison:2261116,Allison:1035669,Agostinelli:602040}-based simulation of a generic collider detector, DICE (Detector In a Collider Experiment). The detector model includes an active calorimeter system and tracker material. 
The calorimeter system comprises an electromagnetic and a hadronic section, inspired by the setup of the CMS experiment up to Run~3. The electromagnetic section consists of a single layer of homogeneous PbWO$_4$ crystals, while the hadronic section is a sampling calorimeter that alternates brass absorbers with polyvinyl toluene sensors. No electronics response or digitisation is simulated; calorimeter energy deposits are used directly. 
In the barrel, the electromagnetic and hadronic section have a segmentation in $\eta \times \phi$ of $256 \times 256$ and $128 \times 128$ respectively, resulting in cell dimensions between $2 \times 2$ cm$^2$ and $7 \times 7$ cm$^2$.
The cells in the endcaps are structured in 40 rings with equal spacing in $\eta$ per layer, with 256 sensors per ring in the ECAL and 128 per ring in the HCAL, resulting in sensor sizes from $3 \times 3$ cm$^2$ to $6 \times 6$ cm$^2$.
The entire detector is immersed in a 4 T magnetic field directed along the beam direction.

This abstraction is sufficient for the present study, which focuses on target-definition effects from shower overlap and detector granularity rather than on readout-specific effects.
The tracker, built out of silicon layers, only serves to model realistic interactions before the calorimeter and is therefore kept inactive.

From the reconstruction point of view, calorimeter hits are built by summing the deposited energy from all particles that traverse the detector cells. Cells with a deposited energy below 10 MeV are removed.

Since tracker hits are not available, the reconstruction uses track proxies to emulate PF-like charged-particle constraints. These proxies are constructed by propagating the charged primary particles from the interaction point as ideal helices in the magnetic field.

This approximation is only meaningful if the particle history originating from the charged primary remains sufficiently collinear while traversing the tracker material. In particular, material interactions can produce secondary particles or large deflections before the calorimeter, in which case a single helix proxy would no longer provide a consistent charged-particle constraint. To reject such cases, a track-quality criterion is introduced.
Tracks are selected by comparing the $p_\text{T}$-weighted distance between the helix position extrapolated to the calorimeter boundary and the calorimeter-boundary positions of all particles originating from the same charged primary particle.

\begin{equation}
    \delta_i = \frac{1}{\sigma}\frac{\sum_j p_{\text{T},j}\sqrt{d_{xy,j}^2 + d_{z,j}^2}}{\sum_j p_{{\text{T},j}}},
\end{equation}

where $\sigma \approx 1$ cm. $d_{xy}$ and $d_z$ are the distances between the positions of the tracks that stemmed from the $i$-th charged mother particle and the position of the helix at the calorimeter boundary.
Good reconstructed tracks are defined as those tracks with $\delta_i < 3$.
Those tracks are represented by a single pseudo-hit placed at the boundary to the calorimeter where the helix enters the calorimeter volume. The reconstructed track momentum magnitude is constructed by applying a relative Gaussian smearing of 1\% to the helix momentum as:
\begin{equation}
    p_{\mathrm{reco}} = p_{\mathrm{helix}}\left(1 + 0.01\,\epsilon\right),
    \qquad \epsilon \sim \mathcal{N}(0,1).
\end{equation}

\section{Truth definition at the boundary}

During the simulation, particles are tracked individually, allowing the identification of those that enter the calorimeter together with the secondary particles they produce. For each particle crossing the calorimeter boundary, all subsequent energy depositions originating from it are associated through the simulation history. This defines a tree of contributions, with the entering particle as the root.

For the purpose of this study, the simulation-history tree is collapsed into a single object by assigning each calorimeter energy deposit, on a hit-by-hit basis, to the particle that entered the calorimeter and forms the root of that history tree. We refer to the resulting object as a \textit{SimShower}. A SimShower therefore represents the full calorimeter response induced by one particle crossing the calorimeter boundary, including the contributions of its secondaries. Fig.~\ref{fig:shower_overlap} illustrates this construction for two pions and the corresponding color-coded SimShowers generated inside the calorimeter.

\noindent
\begin{minipage}{\columnwidth}
    \centering
    \includegraphics[scale=0.5]{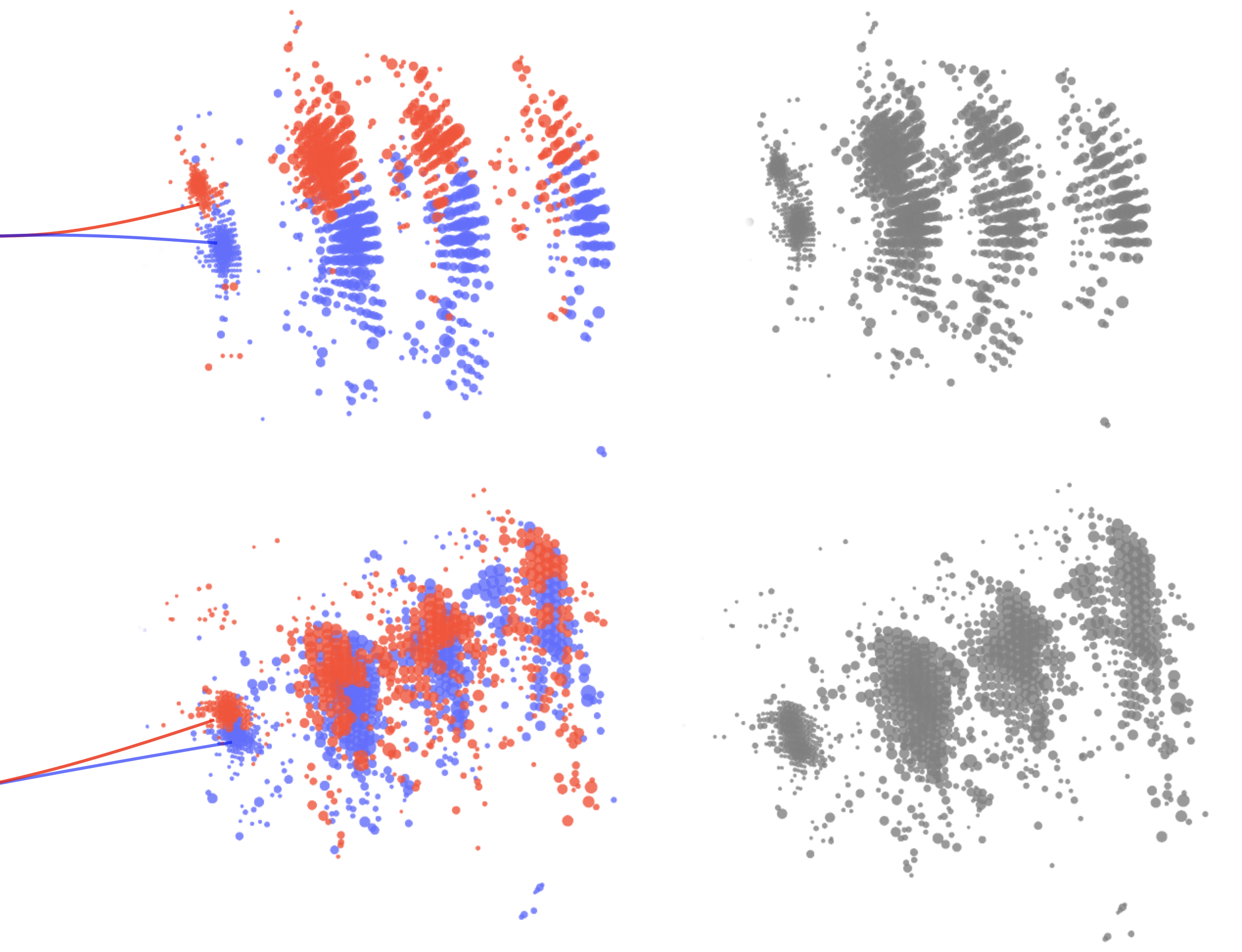}
    \captionsetup{width=.85\columnwidth}
    \captionof{figure}{Illustration of two 100 GeV pions showering in a calorimeter. Top: two particles entering the calorimeter with sufficient separation produce distinct showers that remain experimentally distinguishable. Bottom: two nearby particles lead to strongly overlapping energy deposits that cannot be resolved into separate showers based on the detector response alone.}
    \label{fig:shower_overlap}
\end{minipage}

For each SimShower, several truth and detector-response properties are defined. The \textit{boundary position} is the position of the particle that initiates the SimShower when it crosses into the calorimeter volume. The \textit{boundary energy} and \textit{boundary momentum} are the energy and momentum of the same particle at this point, respectively. These boundary quantities are therefore not derived from the calorimeter hit pattern, but define the kinematic truth information attached to the SimShower. Separately, the \textit{deposited energy} of the SimShower is defined as the sum of all calorimeter energy deposits associated with it through the simulation history. When multiple SimShowers contribute to the same calorimeter cell, their relative contributions are described by cell-wise energy fractions.

While SimShowers provide a well-defined decomposition of the calorimeter response at simulation level, they do not yet account for detector-induced non-resolvability, as illustrated in Fig.~\ref{fig:shower_overlap}.

\section{Shower merging algorithm}

While SimShowers provide a simulation-level decomposition of the calorimeter response, the finite resolving power of the detector is reflected only implicitly through the pattern of shared calorimeter cells. Two nearby particles can produce overlapping energy deposits, leading to SimShowers that contribute to the same cells with non-trivial energy fractions. If such showers are treated as separate reconstruction targets despite being experimentally indistinguishable at the calorimeter granularity, the target definition contains an ambiguity that cannot be resolved from the detector response alone. To make this constraint explicit, SimShowers are processed through a hit-based merging algorithm.

The first step of the algorithm is the assignment of a \textit{resolvability score} to each SimShower. This score quantifies how distinctly a shower contributes to the calorimeter cells it occupies relative to other showers. Shower resolvability is defined as the weighted average of the cell energy fractions\footnotemark[1] attributed to the shower under consideration, as shown in Eq.~\ref{eq:resolvability}.

\begin{equation}
\label{eq:resolvability}
    s_R^A
    =
    \frac{1}{E_A}
    \sum_{j=1}^{N_A} f_j^A E_j^A
    =
    \frac{1}{\sum_{j=1}^{N_A} f_j^A E_j}
    \sum_{j=1}^{N_A} (f_j^A)^2 E_j .
\end{equation}

For a given SimShower $A$:

\begin{itemize}

    \item $s_R^A$ is its resolvability score;

    \item $E_A$ is the total deposited energy associated with the SimShower;

    \item $N_A$ is the number of calorimeter cells to which the SimShower contributes;

    \item $E_j$ is the total deposited energy in cell $j$;

    \item $f_j^A$ is the fraction of $E_j$ attributed to SimShower $A$;

    \item $E_j^A = f_j^A E_j$ is the energy deposited by SimShower $A$ in cell $j$.

\end{itemize}

The connection score $s_C(A \leftarrow B)$ quantifies how strongly SimShower $B$ overlaps with SimShower $A$, normalized to the total deposited energy of $B$. It is defined as

\begin{equation}
    \label{eq:connection}
    \begin{split}
    s_C(A \leftarrow B)
    &=
    \frac{1}{E_B}
    \sum_{j \in (A \cup B)} E_j f_j^A f_j^B  \\
    &=
    \frac{1}{E_B}
    \sum_{j \in (A \cap B)} E_j f_j^A f_j^B .
    \end{split}
\end{equation}

Here, $E_B$ is the total deposited energy associated with SimShower $B$. The second equality follows because at least one of the two fractions vanishes in cells not shared by both SimShowers. The score is directional: $s_C(A \leftarrow B)$ measures the degree to which $B$ is contained in the calorimeter footprint of $A$, and is therefore used to decide whether $B$ should be absorbed into $A$.

A graph is then constructed with SimShowers as nodes and possible absorption relations as directed edges, weighted by the corresponding connection scores. The edges are sorted according to the resolvability state of the connected SimShowers and their connection score, using the following hierarchy:
edges are classified according to the current resolvability state of the absorbing and absorbed SimShowers:

\begin{itemize}
    \item \textit{non-resolvable into non-resolvable}: edges $A \leftarrow B$ for which both $A$ and $B$ are non-resolvable and $s_C(A \leftarrow B)$ is above the connection threshold;
    \item \textit{non-resolvable into resolvable}: edges $A \leftarrow B$ for which $A$ is resolvable, $B$ is non-resolvable, and $s_C(A \leftarrow B)$ is above the connection threshold;
    \item \textit{invalid}: edges connecting two resolvable SimShowers, edges in which a resolvable SimShower would be absorbed into a non-resolvable one, or edges with a connection score below the connection threshold.
\end{itemize}

Within each category, edges are processed in descending order of connection score. After each merging step, the deposited energies, cell fractions, resolvability scores, and edge categories are updated. This dynamic ordering preferentially absorbs non-resolvable fragments into compatible nearby showers, while protecting already resolvable structures from being absorbed. It therefore suppresses the promotion of soft or unresolved shower fragments to independent targets without removing experimentally resolvable structure. This behavior is qualitatively aligned with an infrared- and collinear-safe interpretation of the target definition.

When SimShower $B$ is absorbed into SimShower $A$, the properties of the resulting SimShower are updated as follows:

\begin{itemize}
\item \textbf{Boundary energy}: the boundary energies of the two SimShowers are added.
\item \textbf{Boundary momentum}: the boundary momentum vectors of the two SimShowers are added.
\item \textbf{Hits and fractions}: cells populated only by the absorbed SimShower are added to the absorbing SimShower with their corresponding fractions. For shared cells, the fractions are summed.
\item \textbf{Boundary position}: the boundary positions of the two SimShowers are combined using a boundary-energy-weighted average.
\end{itemize}

The merged SimShower is therefore a composite target object: its boundary quantities define the target kinematics, while its hit fractions define the calorimeter footprint used for subsequent resolvability and connection-score calculations.

\footnotetext[1]{This is the fraction of the total cell energy attributed to the corresponding SimShower.}

\subsection{Global and PF-aware merging}

Two variants of the merging procedure are studied: a \textit{global} approach and a \textit{PF-aware} approach.
In the global approach, all SimShowers are processed simultaneously within the same merging graph.
The PF-aware approach uses the simulation history to distinguish SimShowers associated with charged primary particles from those associated with neutral primary particles.

As illustrated in Figure~\ref{fig:history_aware_flow_chart}, the PF-aware approach sends neutral SimShowers directly to the final global merging stage. For charged primary particles, an intermediate merging step is performed first among all SimShowers associated with the same primary particle passed to GEANT4. This covers, for example, the case of bremsstrahlung radiation in which the showers from the photons overlap with those from the electron.

After this preliminary step, the resulting SimShower with the largest boundary energy, denoted as the \textit{leading shower} in the figure, is tested for compatibility with the corresponding charged primary particle. If the magnitude of the boundary momentum of the leading shower reaches at least 95\% of the primary-particle momentum magnitude, the leading shower is considered a charged Particle-Flow truth target. The SimShowers contained in this leading object are then excluded from the final global merging stage. All remaining SimShowers, including subleading SimShowers associated with the charged primary particle, are passed to the final global merging stage together with the SimShowers associated with neutral primaries and form neutral Particle-Flow truth targets. If the leading shower does not satisfy the 95\% compatibility requirement, no charged Particle-Flow truth target is formed for this primary particle, and all resulting SimShowers from the preliminary step are passed to the final global merging stage.

The PF-aware approach also allows a consistent assignment of the PDG ID for charged-particle candidates through the associated primary particle. In the global approach, such an assignment can become ambiguous because merged SimShowers may originate from different particles. In this case, the PDG ID of the absorbing SimShower is assigned.

\noindent
\begin{minipage}{\columnwidth}
    \noindent
    \centering
    \includegraphics[width=0.85\textwidth]{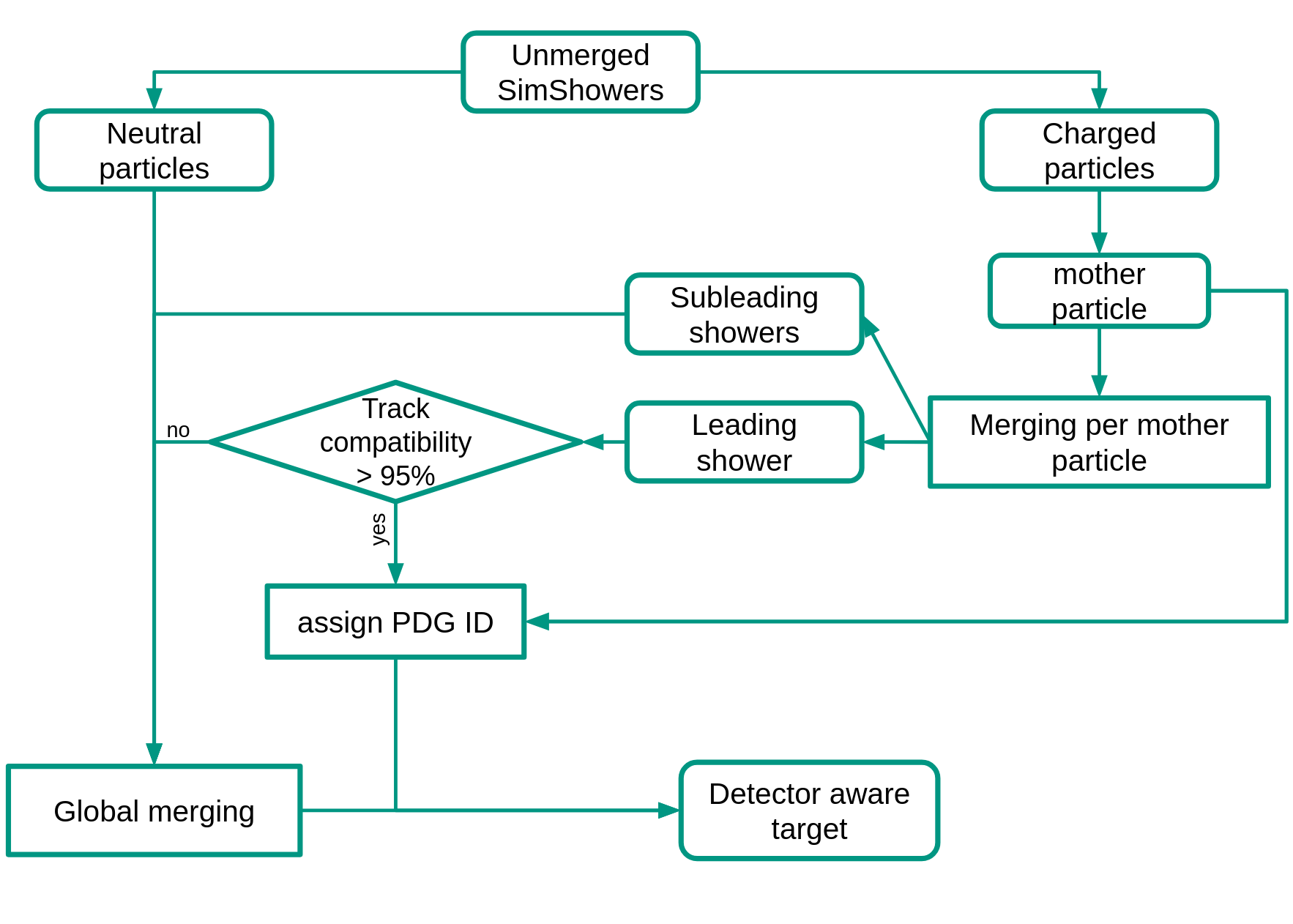}
    \captionsetup{width=.9\columnwidth}
    \captionof{figure}{Flow chart describing the PF-aware merging. The compatibility threshold between the leading shower and the corresponding charged primary particle is set to 95\%.
}
    \label{fig:history_aware_flow_chart}
\end{minipage}

\section{Merging algorithm validation}\label{sec:merging}

The validation of the merging procedure requires a sample with sufficiently dense calorimeter activity to produce a substantial number of overlapping showers, while still allowing a controlled study of the resulting target structures. For this purpose, a ``jet library'' is constructed from $pp \rightarrow t\bar{t}$ events generated with Pythia~8.306~\cite{bierlich2022comprehensiveguidephysicsusage} at a center-of-mass energy of 14~TeV. Stable generated particles are clustered into jets using the anti-$k_\mathrm{T}$ algorithm implemented in FastJet~\cite{Cacciari_2008, FastJet1, FastJet2}. The selected jet constituents of a single jet are subsequently injected as primary particles into the GEANT4 detector simulation to form one event. This single-jet setup is not intended to represent a full physical collision event. Instead, it provides a controlled laboratory for studying shower overlap, target merging, and reconstruction effects in isolation. It also allows specific kinematic regions of the jet library to be selected without introducing additional event-level correlations. For simplicity, we refer to these jets as top-quark jets in the following.

To study the impact of the merging procedure on the resulting target definition, different merging configurations are evaluated through variations of the resolvability threshold. This threshold is the physically relevant parameter of the algorithm, since it determines when a SimShower is considered experimentally distinguishable from its surroundings. The connection threshold, in contrast, is used mainly as a practical graph-pruning criterion: it suppresses absorption candidates with negligible overlap and thereby reduces the combinatorial complexity of the procedure. It should therefore be understood as a technical convenience rather than as an independent physics parameter of the target definition.

\noindent
\begin{minipage}{\columnwidth}
    \centering
    \includegraphics[scale=0.175]{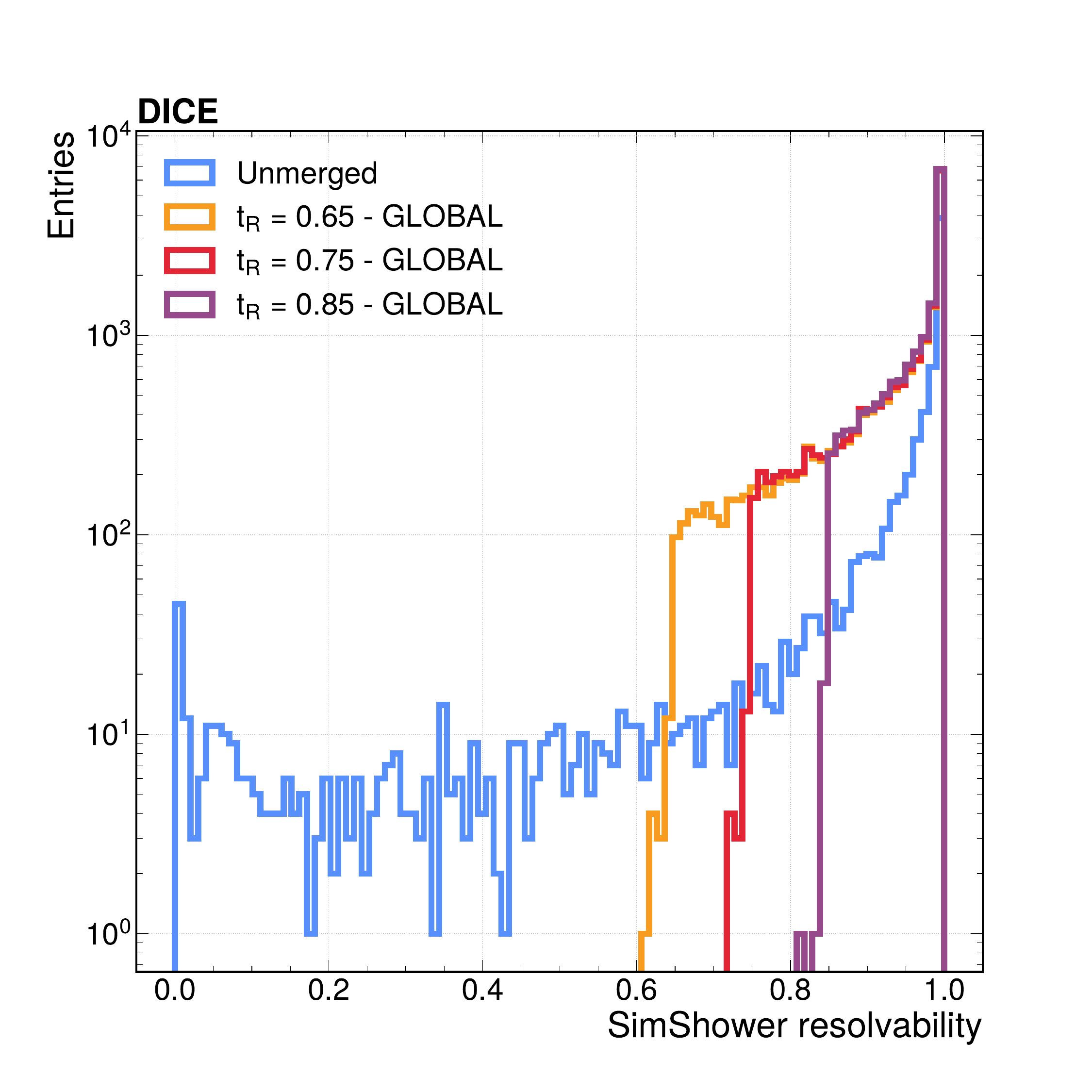}
    \captionsetup{width=.9\columnwidth}
    
    \captionof{figure}{Distribution of the SimShowers' resolvability scores for the unmerged sample and samples merged with different configurations. Higher resolvability scores correspond to topologically more separated SimShowers with a lower amount of shared energy. The effect of the connection score is reflected in the absence of a hard cut-off at the resolvability threshold. }
    \label{fig:sim_shower_resolvability} 
\end{minipage}

Figure~\ref{fig:sim_shower_energy_barendc} shows the SimShower energy distributions for different merging configurations. The connection threshold is fixed to $10^{-4}$, while the resolvability threshold is varied. Increasing the resolvability threshold makes the merging procedure more aggressive, shifting the target-object spectrum toward higher energies. The effect is stronger in the endcap region than in the barrel, consistent with the higher particle density and stronger collimation of forward jets. Consequently, the endcap distributions show a more pronounced reduction of low-energy SimShowers after merging.

\noindent
\begin{minipage}{\columnwidth}
    \centering
    \includegraphics[scale=0.175]{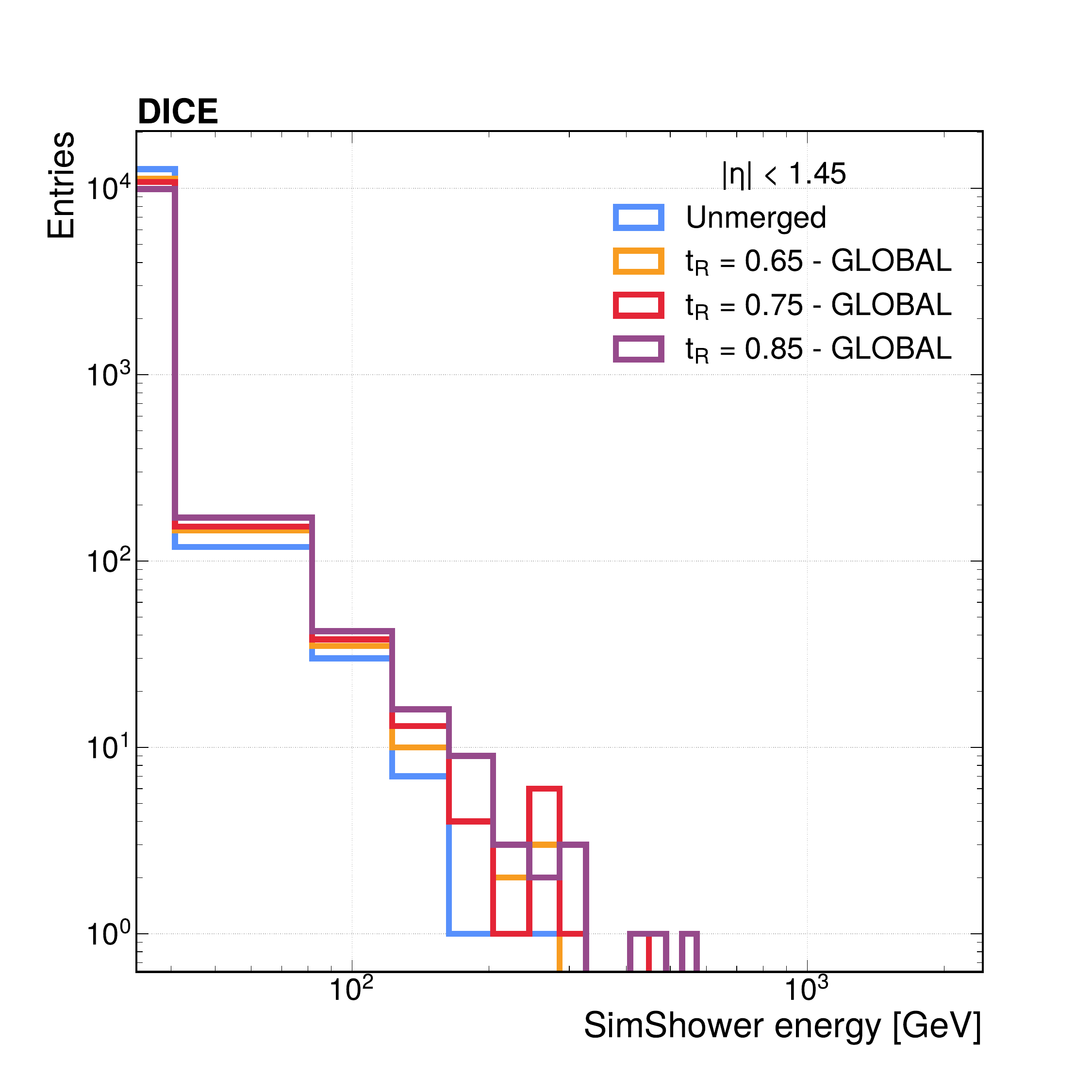}
    \includegraphics[scale=0.175]{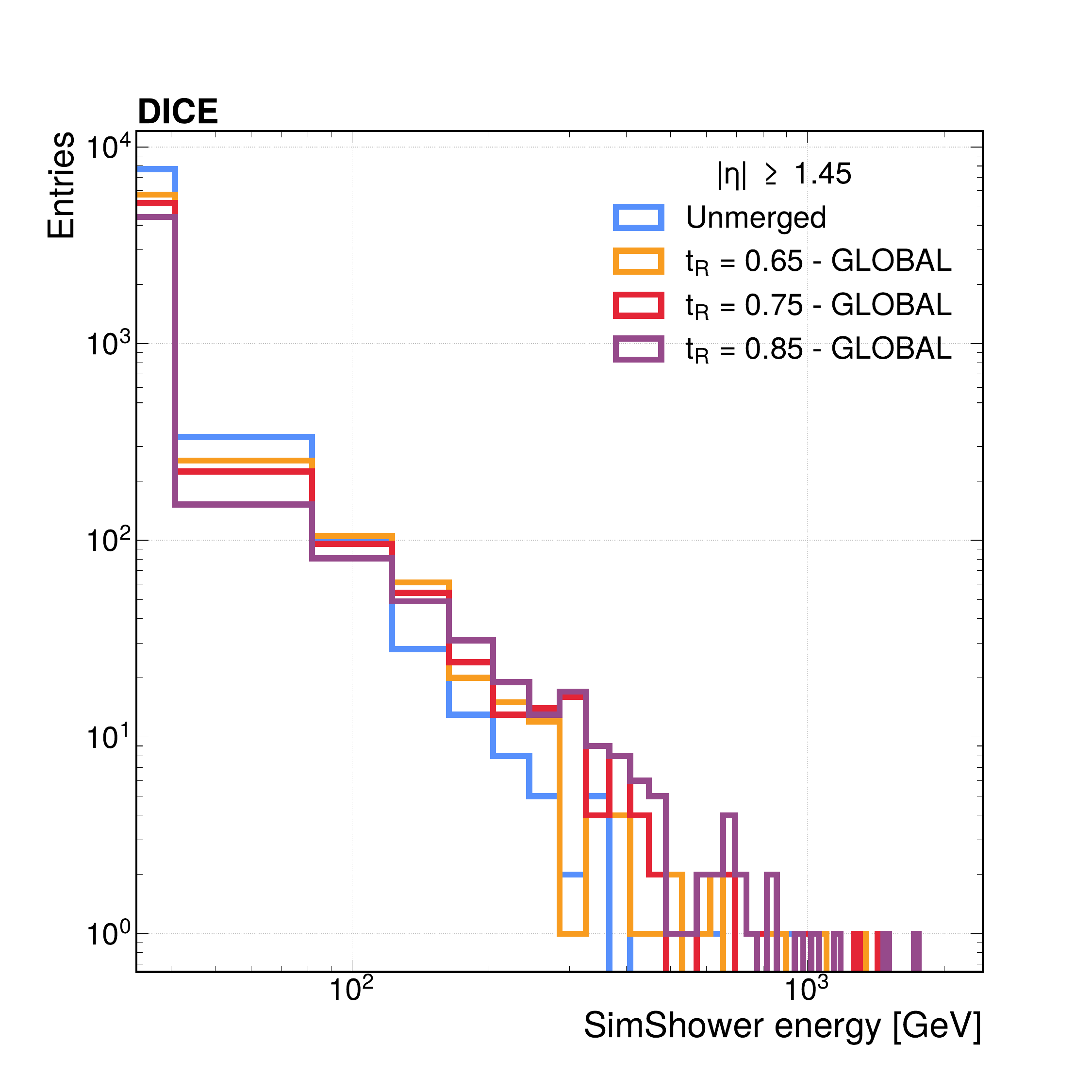}
    \captionsetup{width=.9\columnwidth}
    \captionof{figure}{SimShower energy distributions in the barrel (above) and in the endcaps (below).} 
    \label{fig:sim_shower_energy_barendc} 
\end{minipage}

A complementary observable is the distance of each SimShower to its nearest neighboring SimShower. This probes whether the merging procedure preferentially removes spatially overlapping target objects. The distance is computed at the calorimeter boundary, using the boundary position of each SimShower, and expressed in units of calorimeter cells. Figure~\ref{fig:sim_shower_min_distance_endcap} shows the resulting distribution in the endcap region. As expected, more aggressive merging configurations shift the distribution toward larger nearest-neighbor separations, reflecting the preferential removal of unresolved overlapping structures.

\begin{minipage}{\columnwidth}
    \noindent
    \centering
    \includegraphics[scale=0.175]{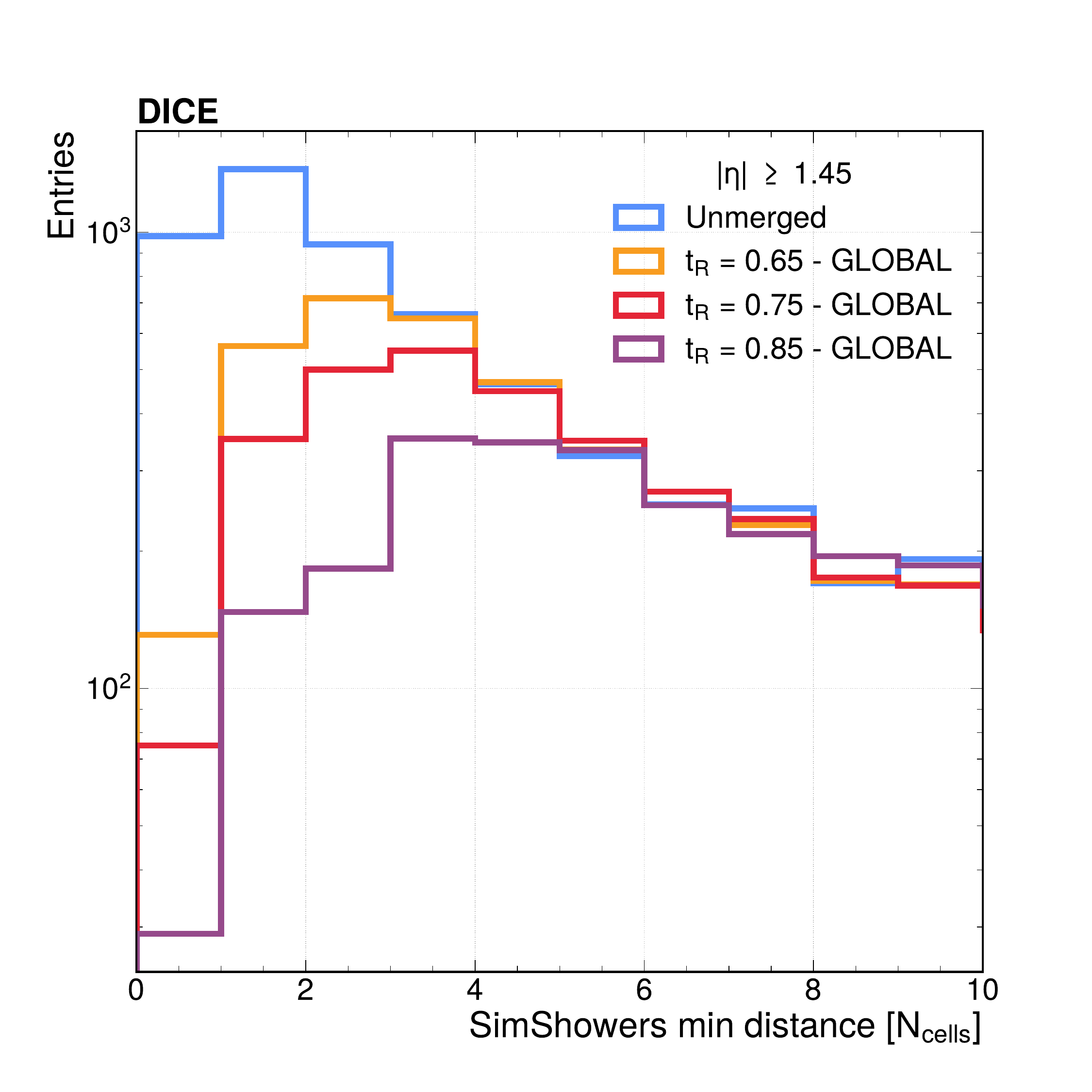}
    \captionsetup{width=.9\columnwidth}
    \captionof{figure}{Distance of each SimShower to its nearest neighboring SimShower in the endcap region, computed from the boundary positions and expressed in units of calorimeter cells.}
    \label{fig:sim_shower_min_distance_endcap}
\end{minipage}

The PF-aware merging systematically produces milder modifications of the target structure than the global approach for the same set of thresholds, as illustrated in Fig.~\ref{fig:sim_shower_energy_barendc_diff}. In the PF-aware variant, charged-particle-related SimShowers are first processed within separate particle-associated groups before entering the final global stage. As a consequence, overlapping SimShowers originating from different charged particles are protected from being merged during this intermediate step. This preserves more fine-grained calorimeter structures and results in a less pronounced shift of the SimShower energy spectrum toward high energies.

\begin{minipage}{\columnwidth}
    \noindent
    \centering
    \includegraphics[scale=0.175]{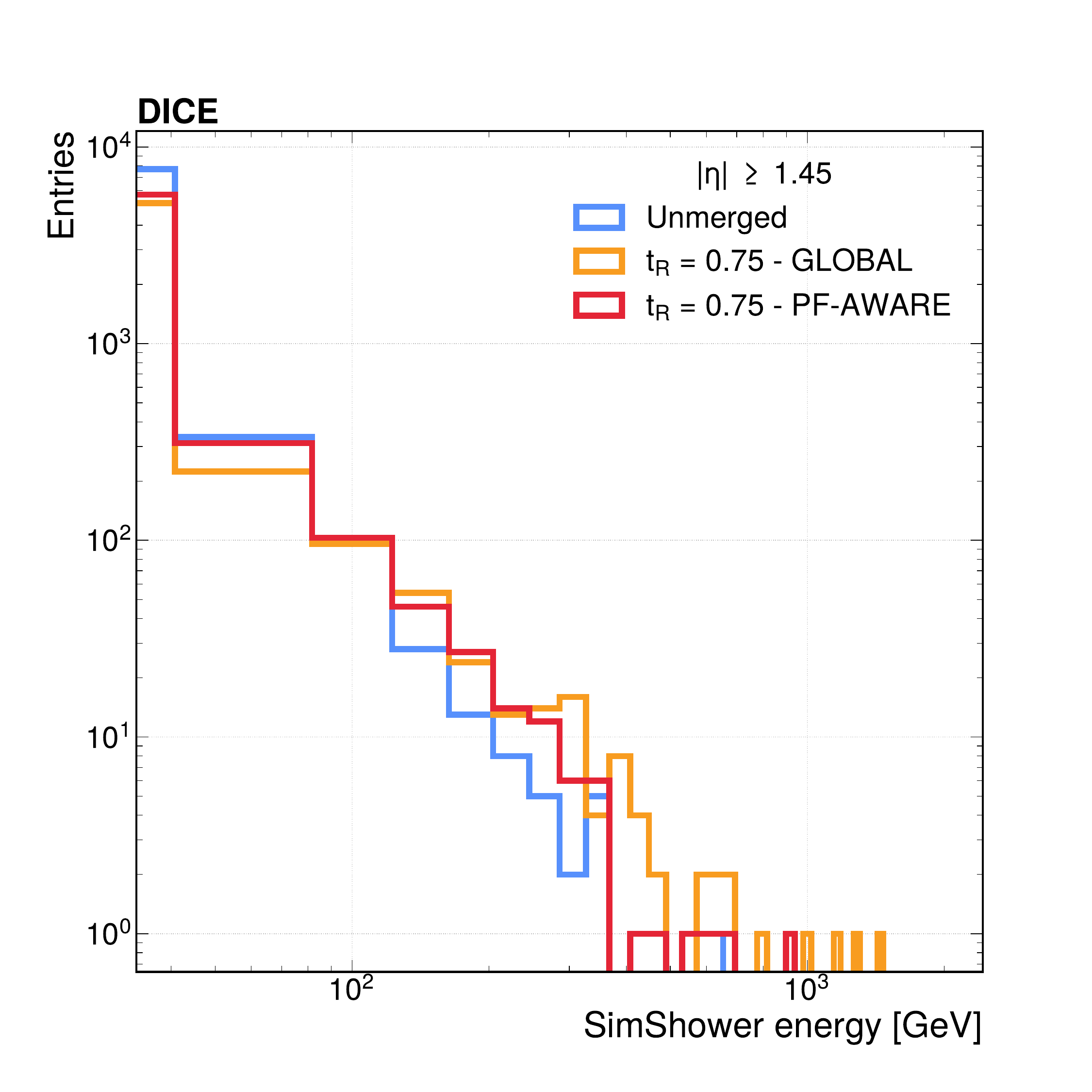}
    \captionsetup{width=.9\columnwidth}
    \captionof{figure}{SimShower energy distributions in the endcap region for the unmerged, globally merged, and PF-aware merged samples. For identical thresholds, the global merging produces a stronger shift toward high-energy SimShowers, reflecting the more aggressive absorption of overlapping structures.}
    \label{fig:sim_shower_energy_barendc_diff}
\end{minipage}

\section{Training setup and same-process results}

To study the impact of the target definition on reconstruction performance, a ML-based Particle Flow reconstruction model is trained using the different merging configurations introduced in the previous sections. By keeping the model architecture fixed, this study isolates the effect of the target definition on the resulting reconstruction performance and robustness. The underlying assumption is that, for a sufficiently expressive reconstruction model, ambiguities inherent in the target definition become a limiting factor for the achievable reconstruction quality.

For this purpose, a GNN-based reconstruction model using GravNet layers~\cite{GravNet} and trained with the Object Condensation (OC) method~\cite{ObjCond} is employed. The full architecture is shown in Figure~\ref{fig:merging_model} and follows previous studies targeting calorimeter clustering and Particle Flow~\cite{E2EReco_HighOcc, HGCAL_MultiparticleReco, PF_GNNs_FCC}. 

The input consists of calorimeter hits and pseudo-hits representing tracks. For each input object, the energy for calorimeter hits or momentum magnitude for track pseudo-hits, the charge, and positional information are passed to the neural network. Calorimeter hits are assigned zero charge.

The model is trained on a sample of 100\,000 jets from the jet-library dataset described in Sec.~\ref{sec:merging}, consisting of $\approx 500$ hits per event. Jets are selected with $p_\mathrm{T} > 30~\mathrm{GeV}$ and either $|\eta| < 1.2$ or $1.7 < |\eta| < 3.0$, corresponding to barrel-like and endcap-like topologies. SimShowers are assigned to the reconstruction target if their calorimeter boundary position falls within $|\eta| < 1.4$ or $1.55 \leq |\eta| \leq 3.0$. SimShowers with boundary energies below $0.1~\mathrm{GeV}$ are treated as noise.
The training has been performed on a NVIDIA A100 for 300 epochs, using batches consisting of up to 50\,000 hits.

The impact of the target definition is evaluated for the unmerged case and for several merged configurations. Unless stated otherwise, the connection threshold is fixed to $t_C = 10^{-4}$, while the resolvability threshold $t_R$ is varied. This choice follows the interpretation discussed in Sec.~\ref{sec:merging}: $t_R$ controls the physically relevant degree of target merging, whereas $t_C$ is used as a graph-pruning criterion.
All performance plots are evaluated on statistically independent samples not used for training.

\begin{minipage}{\columnwidth}
    \noindent
    \centering
    \includegraphics[scale=0.25]{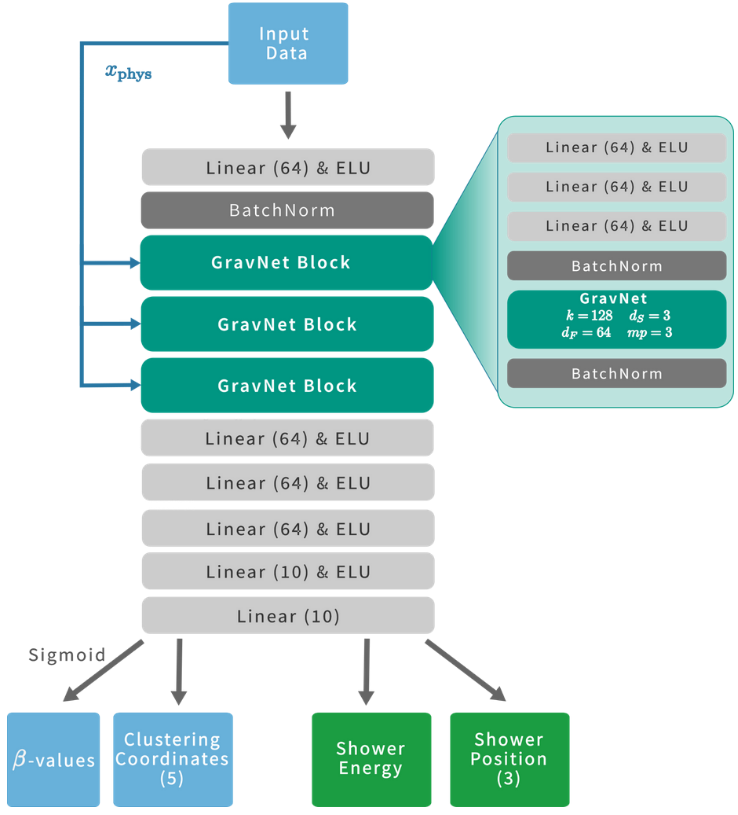}
    \captionsetup{width=.9\columnwidth}
    \captionof{figure}{Schematic representation of the ML-based Particle Flow reconstruction model used to evaluate the impact of different target definitions.}
    \label{fig:merging_model}
\end{minipage}

Since the merging procedure changes the shower-level target definition, particle-by-particle comparisons are not well defined across different merging configurations. The reconstruction performance is therefore evaluated using anti-$k_T$ jet observables, which provide an infrared- and collinear-safe reference at particle level and are therefore expected to be largely insensitive to the redistribution of unresolved target structure. In the present detector-level setup, small residual differences can still occur near the barrel-endcap transition and acceptance boundaries, where changes in the merged-object barycenter may move objects into or out of the selected detector region. This residual effect is illustrated in Fig.~\ref{fig:jet_diff_pt}. The distribution is strongly peaked around zero for both merging configurations, showing that the merged target definitions preserve the truth-jet transverse momentum to good accuracy. The remaining tails are small compared to the reconstructed jet resolutions studied below.

For the performance evaluation, only jets with $p_\mathrm{T} > 10~\mathrm{GeV}$ are considered. Reconstructed jets are matched to true jets if their separation in the $\eta$--$\phi$ plane satisfies $\Delta R < 0.4$.

\begin{minipage}{\columnwidth}
    \centering
    \includegraphics[scale=0.25]{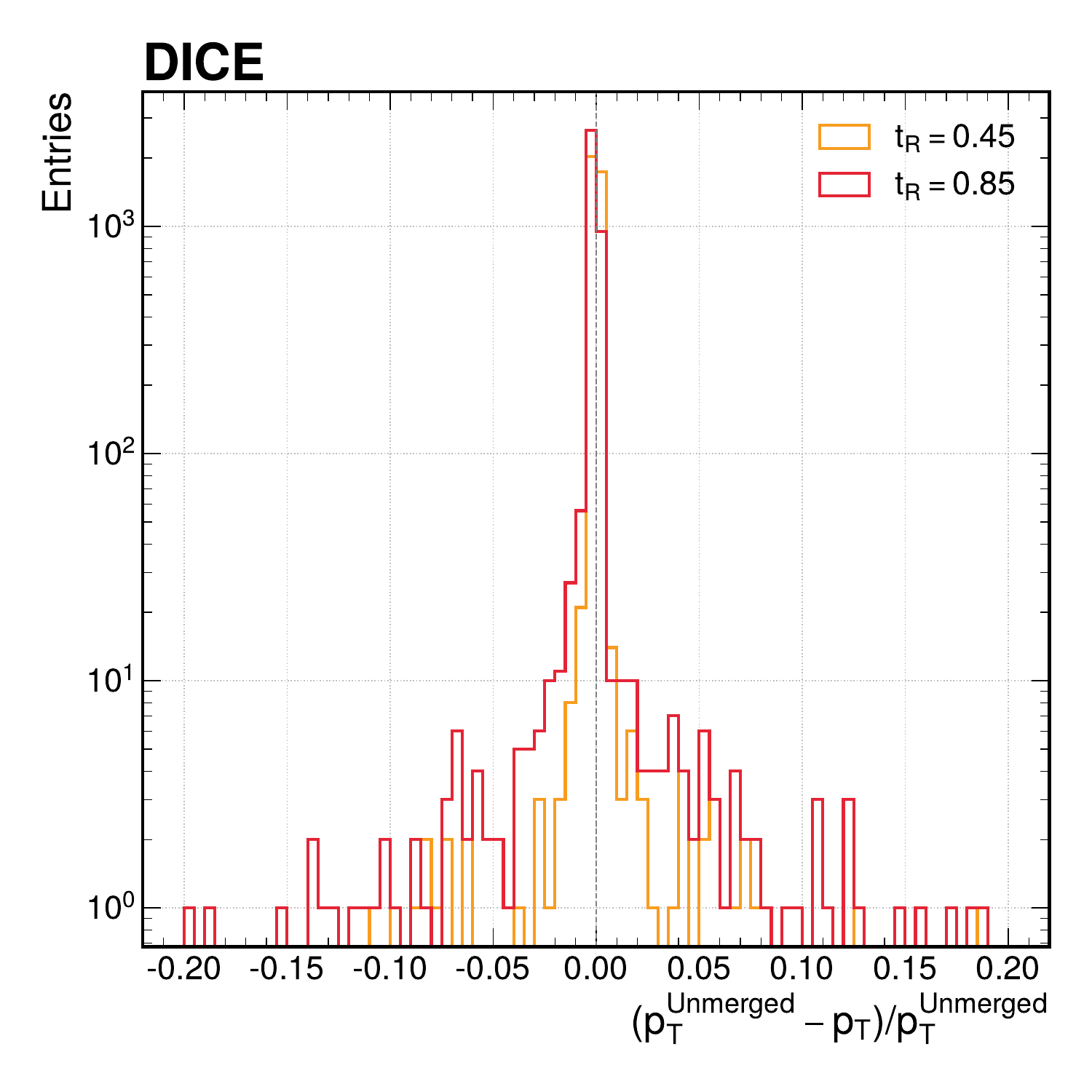}
    \captionsetup{width=.9\columnwidth}
    \captionof{figure}{Relative difference in truth-jet $p_T$ for PF-aware merged target definitions with $t_R = 0.45$ and $t_R = 0.85$ with respect to the unmerged target definition. Differences arise mainly when changes in the merged-object barycenter move objects across vetoed detector regions or acceptance boundaries.}
    \label{fig:jet_diff_pt}
\end{minipage}

\begin{figure*}[t!]

    \centering

    \includegraphics[scale=0.2]{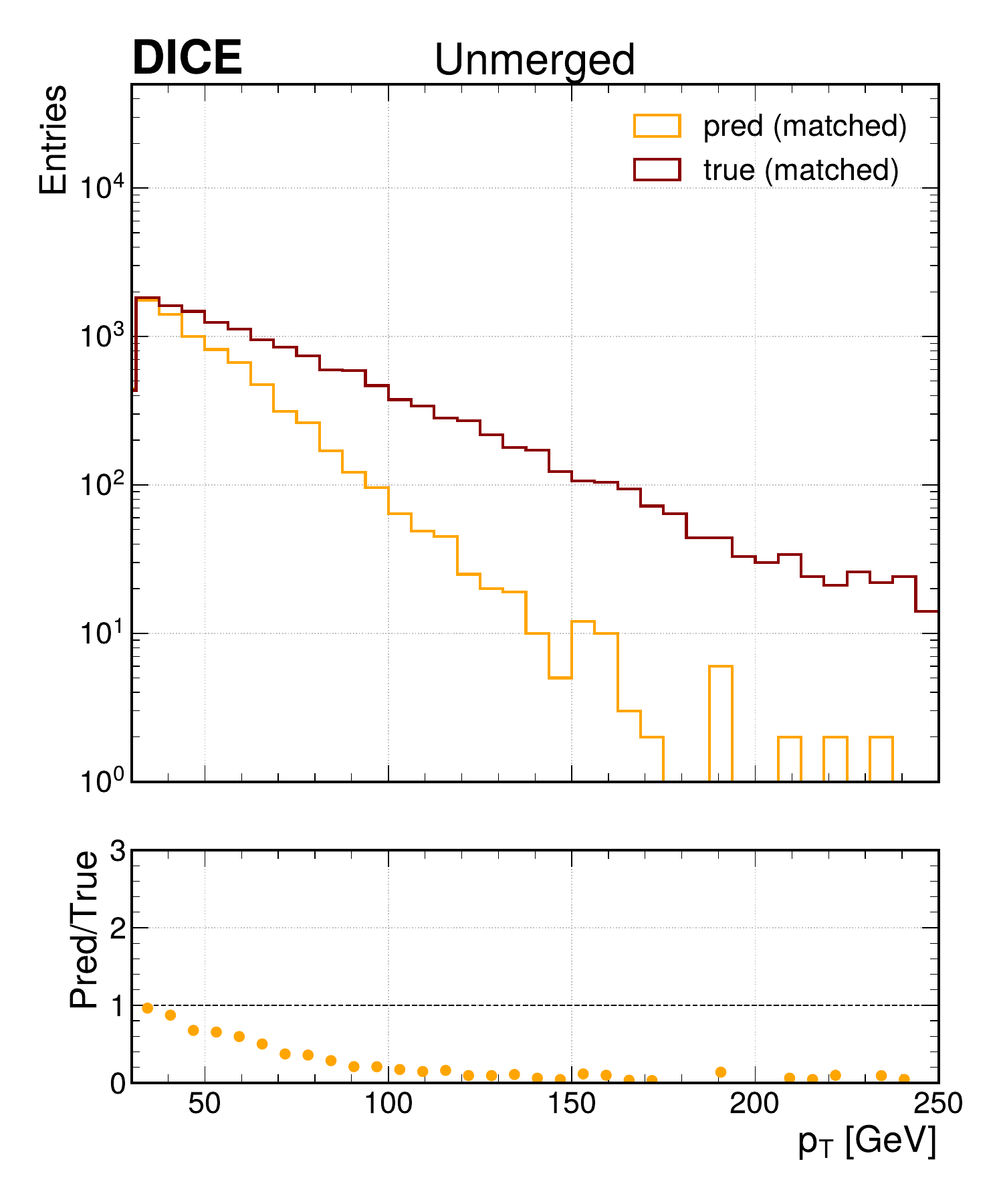}
    \includegraphics[scale=0.2]{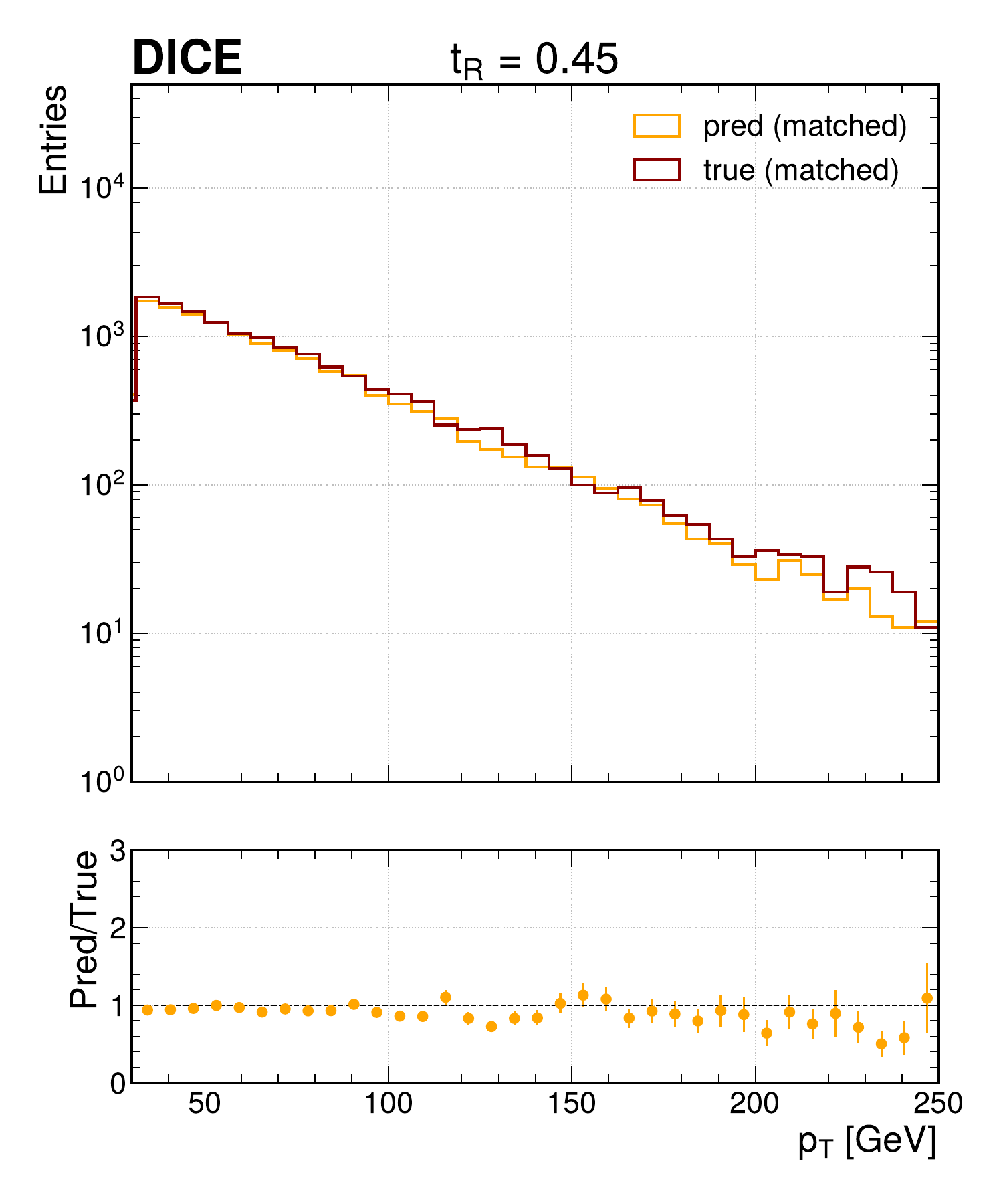}\hspace{5cm}
    \includegraphics[scale=0.2]{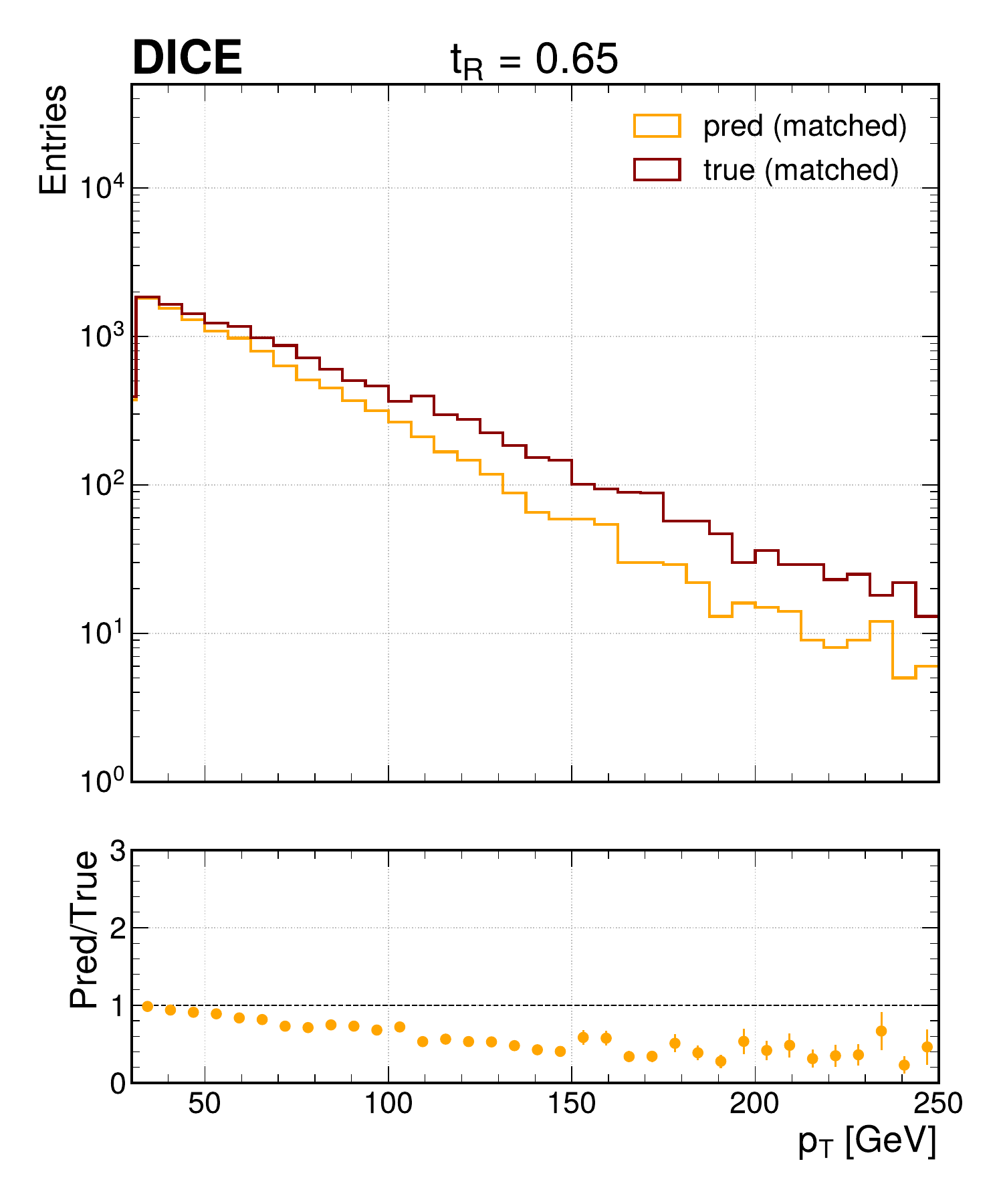}
    \includegraphics[scale=0.2]{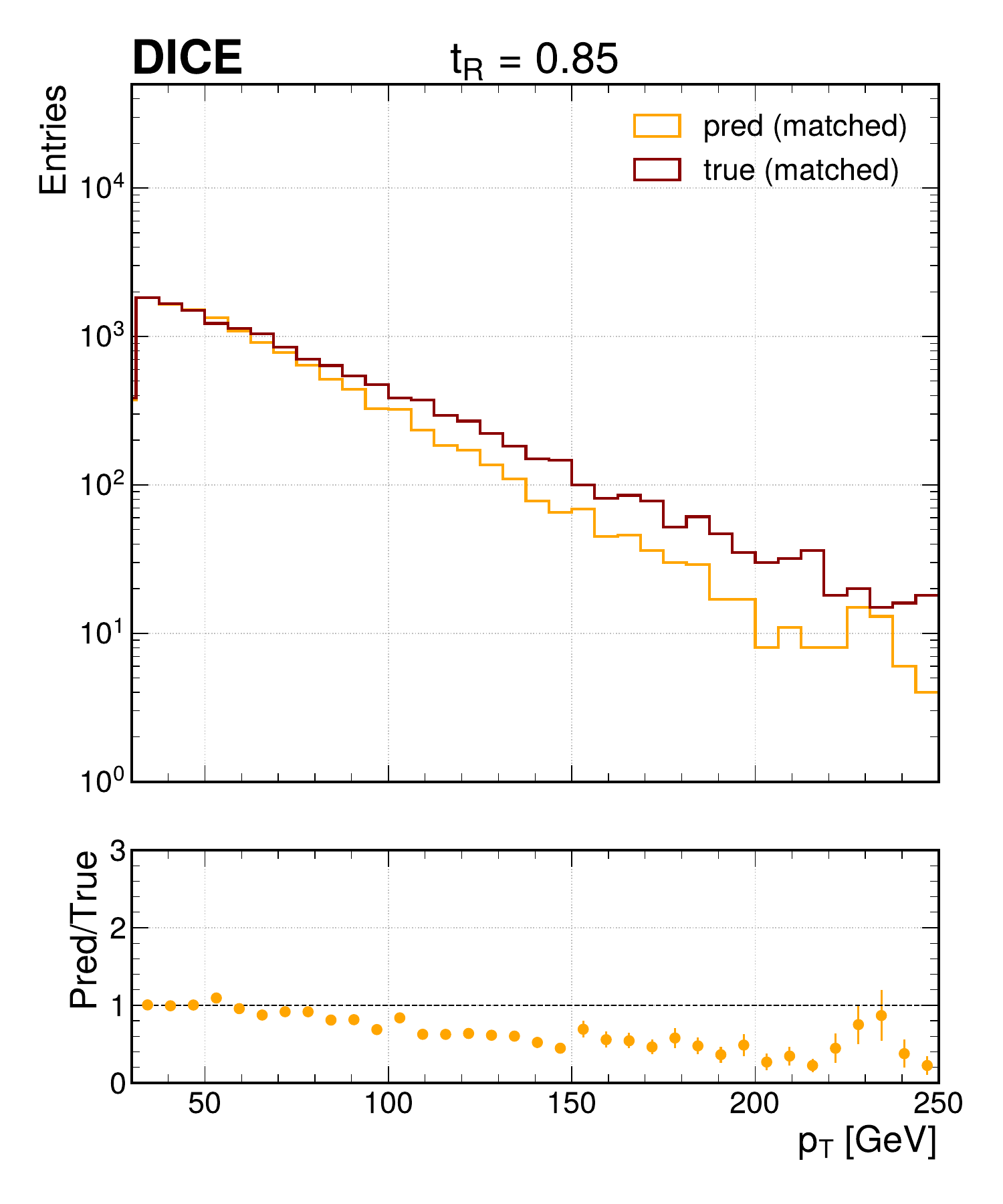}
    \captionsetup{width=0.8\textwidth}

    \caption{Transverse momentum distributions of jets from top quarks. The predicted values are obtained from models that have been trained on different truth information samples: one used unmerged truth, while the other three were trained on PF-aware merged truth with, respectively, $t_R = 0.45$, $t_R = 0.65$, and $t_R = 0.85$.}

    \label{fig:topjets_pt_distribution}

\end{figure*}

By comparing the transverse momentum distributions of reconstructed and true jets from top quarks, Fig.~\ref{fig:topjets_pt_distribution} illustrates the effect of well-defined reconstruction targets. The reconstructed jet $p_\mathrm{T}$ spectrum begins to fall off already below 100~GeV in the unmerged case. In contrast, even a mild PF-aware merging configuration, such as $t_R = 0.45$, exhibits an improved reconstructed transverse momentum spectrum that yields better agreement with the true distribution. This improvement is particularly pronounced for jets with transverse momenta well beyond 200~GeV, where the target redefinition results in a better reconstruction efficiency.

\begin{minipage}{\columnwidth}
    \noindent
    \centering
    \includegraphics[scale=0.25]{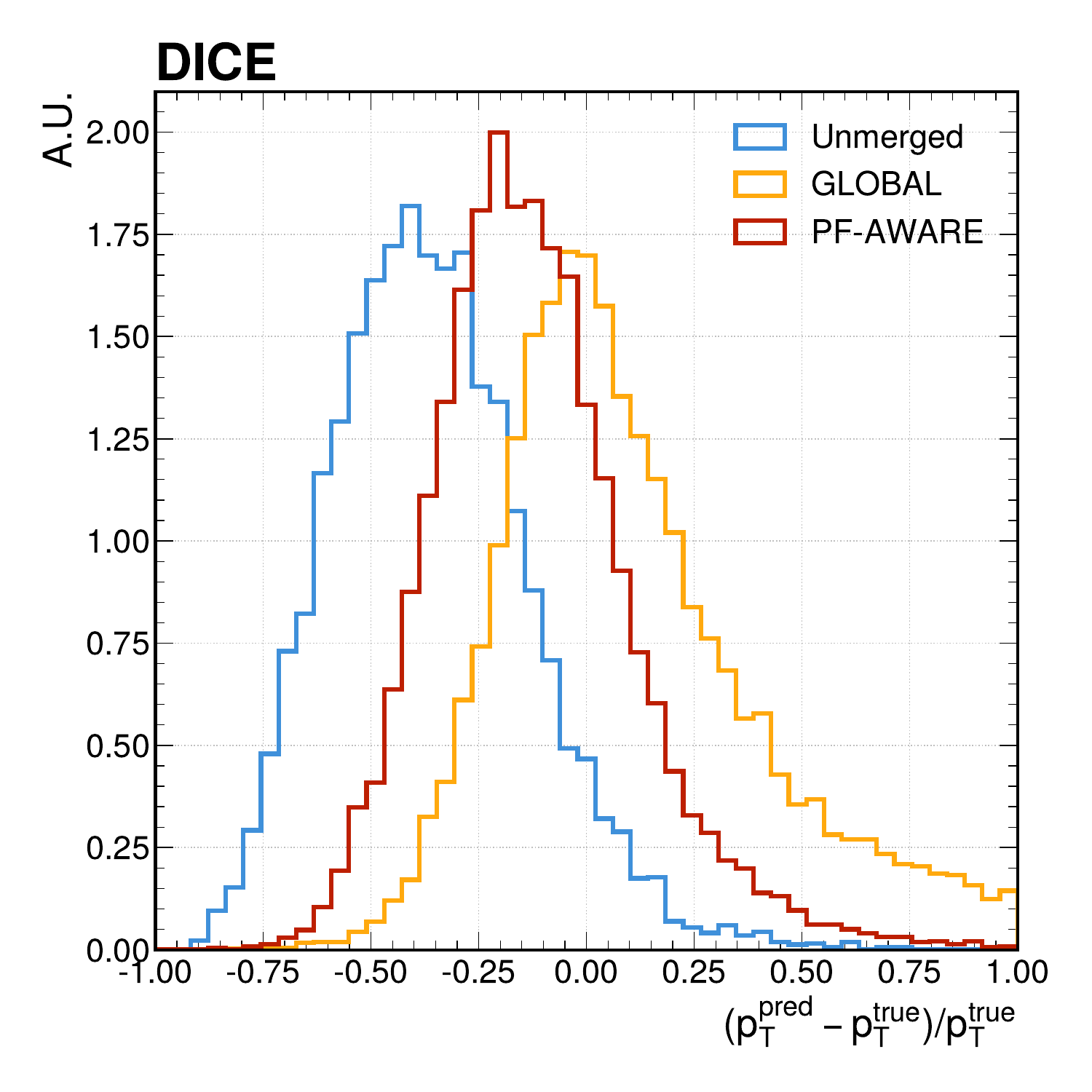}
    \captionsetup{width=.9\columnwidth}
    \captionof{figure}{Transverse momentum response in the top quark sample for models trained with different target definitions: unmerged, globally merged, and PF-aware merged targets. Both merging approaches use a resolvability threshold of $t_R = 0.85$. Although global and PF-aware merging achieve comparable accuracy, the latter one exhibits a higher precision, whereas global merging shows a tendency toward overestimation.}
    \label{fig:topjets_pt_global_iterative}
\end{minipage}

To assess the impact of the two merging strategies, as well as their differences, both global and PF-aware merging are evaluated on the top-quark jet dataset for a relatively strong merging configuration of  $t_R = 0.85$. The connection threshold is kept fixed at $t_C = 10^{-4}$. Figure~\ref{fig:topjets_pt_global_iterative} shows the jet $p_\mathrm{T}$ response for three models: one trained on unmerged targets, one trained on globally merged targets, and one trained on PF-aware merged targets.
The models trained on merged truth show a higher accuracy in the prediction of the transverse momentum, with a tendency towards overestimation for the global approach, which has a notably worse precision than the PF-aware one.

Considering the treatment of charged-particle information, the observed difference between the two merging approaches is to be expected. In the global approach, merged SimShowers may originate from particles with different charges, and the PDG ID and charge assignment are inherited from the absorbing SimShower. If a SimShower originating from a neutral particle absorbs one originating from a charged particle, the corresponding tracker-associated information can be lost, making the downstream Particle-Flow reconstruction less consistent. The PF-aware approach avoids this by first processing SimShowers associated with the same charged particle and preserving the corresponding charged-particle information. This provides a natural explanation for the improved response observed in Fig.~\ref{fig:topjets_pt_global_iterative}. For this reason, the remaining robustness studies use the PF-aware merging approach exclusively.

\section{Cross-process robustness of merged targets}

The central question of this study is whether reducing detector-induced target ambiguity improves the robustness of the reconstruction model outside the topology on which it was trained. If unresolved ambiguities are left in the training labels, the model has to resolve them through correlations present in the training sample. This can introduce a dependence on sample-specific priors and may degrade performance when the model is applied to objects with different particle composition or shower topology.

To test this, models trained on the top-quark jet library described in Sec.~\ref{sec:merging} are evaluated on an independent sample built from single $\tau$ leptons. In this context, a $\tau$ jet is defined operationally as the jet obtained by clustering the visible final-state particles from the $\tau$ decay. The sample includes both hadronic and leptonic $\tau$ decays; consequently, the reconstructed object may range from a narrow hadronic $\tau$ jet to an isolated electron or muon. This provides a controlled change in particle composition and detector topology relative to the top-quark jet library, and tests whether the target definition improves the stability of the reconstruction outside the training sample composition.

The $\tau$-jet sample is generated by injecting single $\tau$ leptons into the detector simulation within the calorimeter acceptance.
The sample consists of 5\,0000 $\tau$-leptons with an exponential $p_\text{T}$ spectrum ranging from 30~GeV up to 300~GeV, restricted to $|\eta| < 1.2$ or $1.7 < |\eta| < 3.0$.

Truth jets are defined by clustering the visible particles from the $\tau$ decay, excluding neutrinos. The same jet selection and matching criteria as for the top-quark sample are then applied to both truth and reconstructed jets.

The relative jet $p_\mathrm{T}$ response is shown in Fig.~\ref{fig:response_all_tau} for two different configurations of the PF-aware merging approach and the unmerged sample. While both models trained on targets merged with the PF-aware approach clearly outperform the model trained on unmerged targets in terms of the central value of the response distribution, the configuration with $t_R=0.45$  additionally exhibits a narrower distribution in comparison with the unmerged case.

\begin{minipage}{\columnwidth}
    \noindent
    \centering
    \includegraphics[scale=0.25]{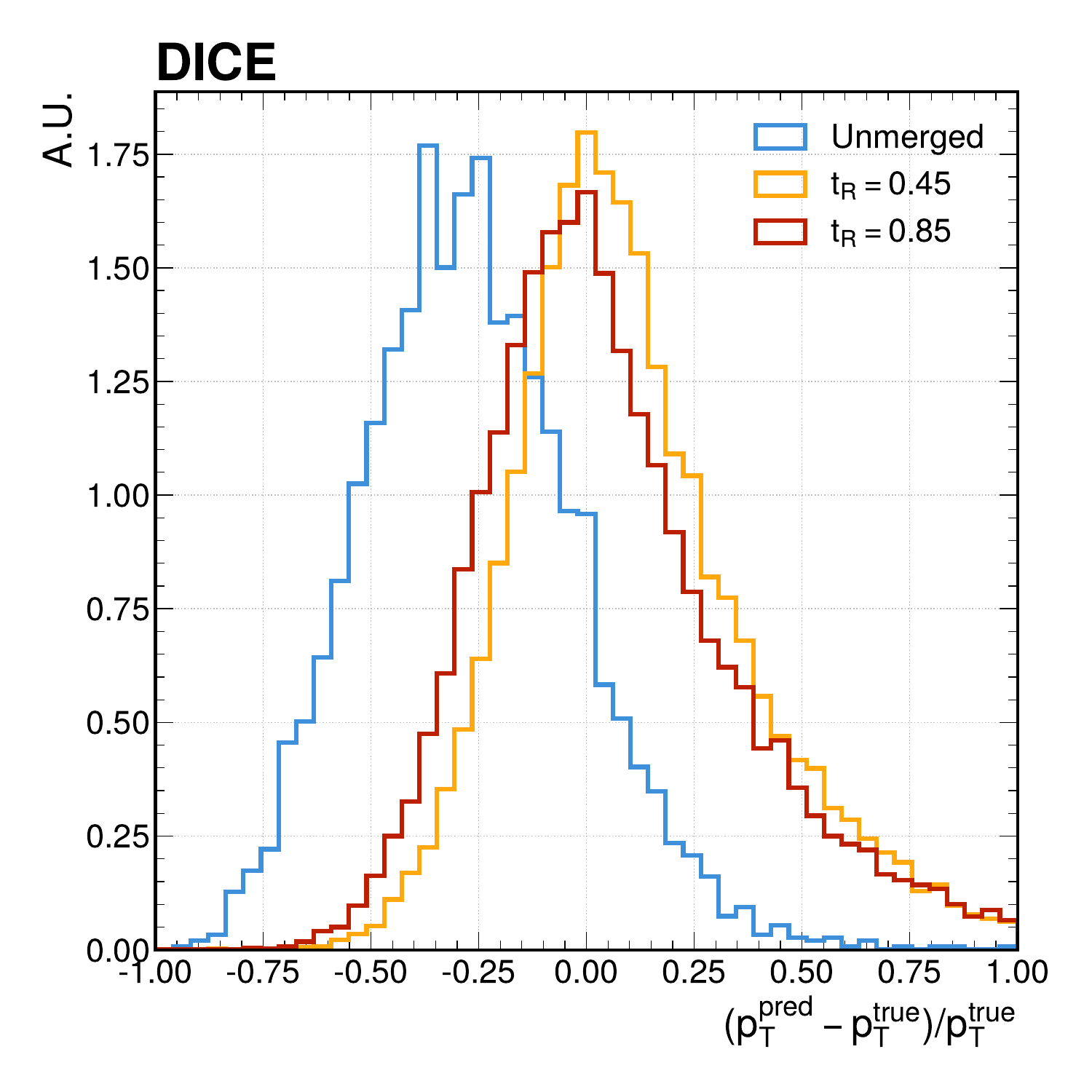}
    \captionsetup{width=.9\columnwidth}
    \captionof{figure}{Relative jet $p_\mathrm{T}$ response in the $\tau$-jet sample. Models trained on PF-aware merged targets show improved central response and narrower response distributions compared to the model trained on unmerged targets.}
    \label{fig:response_all_tau}
\end{minipage}

\begin{figure*}[!t]
    \centering
    \includegraphics[scale = 0.25]{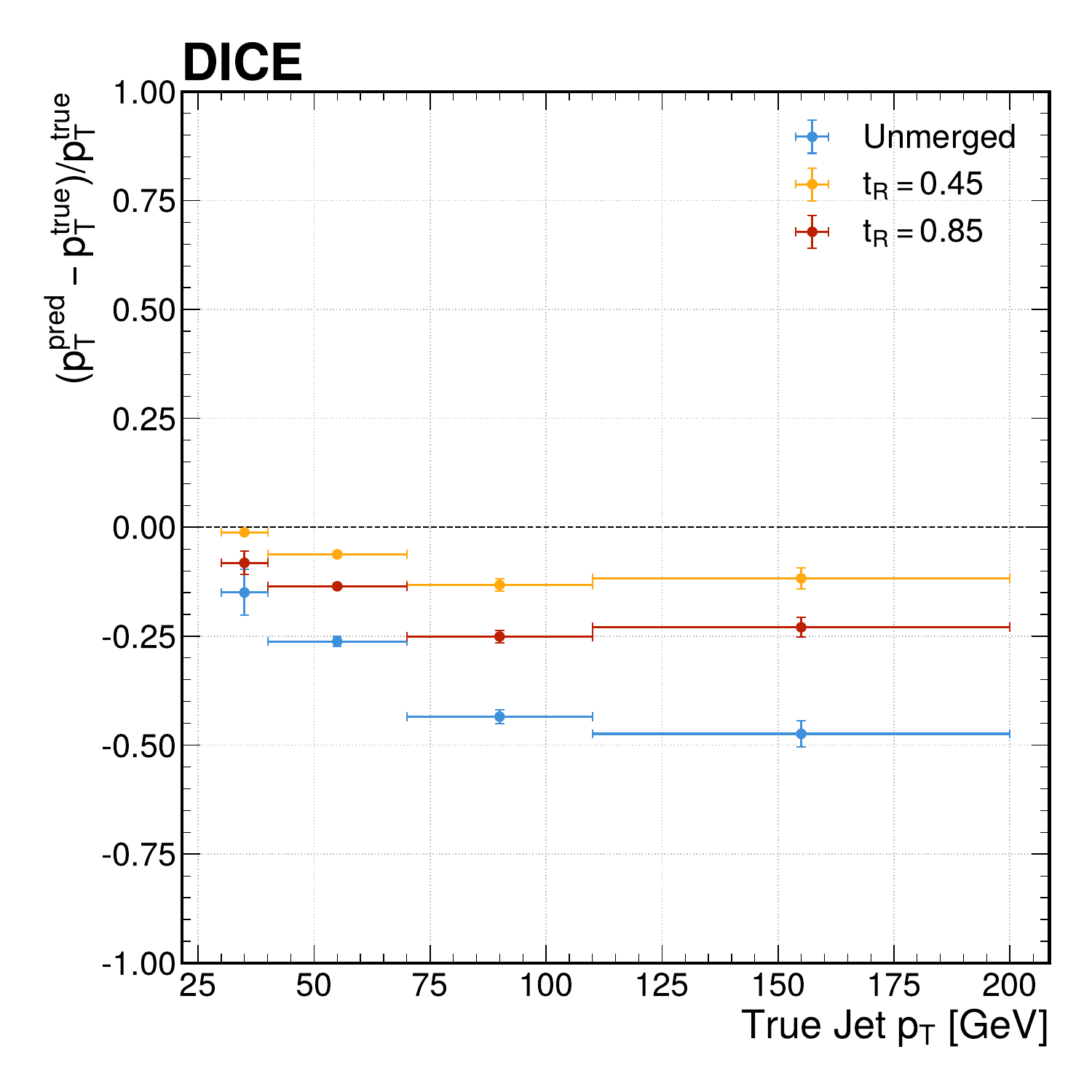}
    \includegraphics[scale = 0.25]{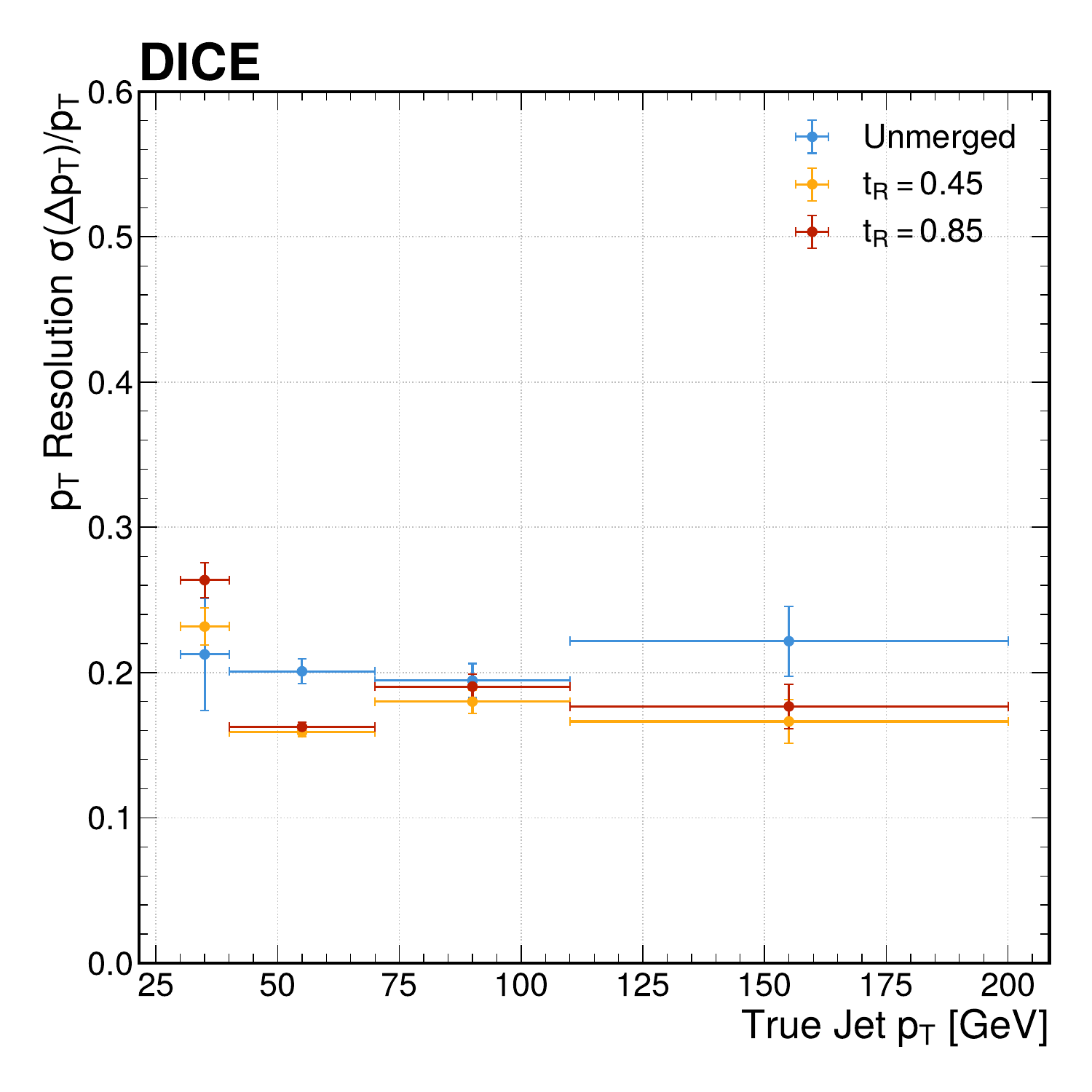}
    \caption{Relative jet $p_\mathrm{T}$ response and resolution as a function of the true jet transverse momentum in the $\tau$-jet sample.}
    \label{fig:response_resolution_tau}
\end{figure*}

Figure~\ref{fig:response_resolution_tau} shows the relative jet $p_\mathrm{T}$ response and resolution as a function of the true jet transverse momentum.
The response and resolution are obtained by fitting the $p_\text{T}$ residuals between the predicted and true jets with a double-sided crystal ball function. 
The response is defined as the central value of the fitted distribution, and the resolution as its width. The error bars represent the uncertainties of the corresponding fit parameters.

Merged truth information yields a better response across the entire momentum range, while resolution is comparable in the lowest $p_\mathrm{T}$ bin, with improvements appearing moving to larger momenta and reaching $\sim 20\%$ between 40 and 70 GeV.

The improved performance on $\tau$ jets, despite training only on top-quark jets, supports the interpretation that detector-aware merging reduces sample-prior dependence in the learned reconstruction. By removing target structures that are not experimentally resolvable, the merged target definition provides a more stable learning objective and improves robustness under changes in jet topology.

\section{Summary}

This work studies the impact of detector-aware reconstruction targets on ML-based Particle Flow reconstruction. In dense calorimeter environments, different simulation-level shower decompositions can lead to detector responses that are experimentally indistinguishable. Treating such non-resolvable structures as separate training targets introduces an ambiguity that cannot be resolved from detector information alone and can force supervised models to rely on sample-dependent priors.

We presented a hit-based SimShower merging procedure that makes detector-induced non-resolvability explicit in the target definition. The algorithm uses cell-wise energy sharing to quantify shower resolvability and merges non-resolvable structures into detector-resolvable target objects. Two variants were studied: a global approach, which processes all SimShowers simultaneously, and a PF-aware approach, which preserves charged-particle consistency before the final global merging stage.

The resulting target definitions were evaluated using a fixed ML-based Particle Flow reconstruction model. On top-quark jets, models trained with merged targets reduce the low-energy bias observed for unmerged shower targets and improve the reconstructed jet transverse-momentum spectrum. The PF-aware merging variant gives the most consistent behavior, because it preserves the charged-particle information needed for Particle Flow reconstruction.

The main physics result is the improved robustness under a change of jet topology. Models trained on top-quark jets and evaluated on independent $\tau$-jet samples show significantly improved jet response and resolution when trained with merged targets. 
This supports the interpretation that detector-aware target merging reduces the dependence of the learned reconstruction on process-specific sample priors. The target definition is therefore not merely a bookkeeping choice: it directly affects the physics performance and robustness of ML-based reconstruction algorithms.

\section{Acknowledgements}
JK and AB were supported by the Alexander von Humboldt Foundation for part of this work. AB received funding from the ErUM-Data BRAID Consortium during the second part of the project.

\FloatBarrier
\newpage    
\printbibliography

\end{multicols}
\end{document}